\documentstyle[aps,twocolumn,prl,epsfig]{revtex}

\newcommand\be{\begin{eqnarray}}
\newcommand\ee{\end{eqnarray}}

\begin{document}
\title{Efficient and robust entanglement generation in a many-particle system
 with
resonant dipole-dipole interactions}
\author{R.G. Unanyan $^{1,2}$ and M. Fleischhauer $^1$}
\address{$^1$ Fachbereich Physik, Universit\"at Kaiserslautern, 67653\\
Kaiserslautern, Germany}
\address{$^2$ Institute for Physical Research of Armenian National\\
Academy of Sciences, Ashtarak-2 378410, Armenia}
\date{\today}
\maketitle

\begin{abstract}
We propose and discuss a scheme for robust and efficient generation of
many-particle entanglement in an ensemble of Rydberg atoms with resonant
dipole-dipole interactions. It is shown that in the limit of complete dipole
blocking, the system is isomorphic to a multimode
Jaynes-Cummings model. While dark-state population transfer is not capable
of creating entanglement, other adiabatic processes are identified that lead
to complex, maximally entangled states, such as the $N$-particle analog of
the GHZ state in a few steps. The process is robust, works for even and odd
particle numbers and the characteristic time for entanglement generation
scales with $N^\alpha$, with $\alpha$ being less than unity.
\end{abstract}

\pacs{42.50-p,32.80.-t}

%%%%%%%%%%%%%%%%%%%%%%%%%%%%%%%%%%%%%%%%%%%%%%%%%%%%%%%%%%%%%%%%%%%%%%%%%%
%%%%%%%%%%%%%%%%%%%%%%%%%%%%%%%%%%%%%%%%%%%%%%%%%%%%%%%%%%%%%%%%%%%%%%%%%%

%%%%%%%%%%%%%%%%%%%%%%%%%%%%%%%%%%%%%%%%%%%%%%%%%%%%%%%%%%%%%%%%%%%%%%%%%%
%%%%%%%%%%%%%%%%%%%%%%%%%%%%%%%%%%%%%%%%%%%%%%%%%%%%%%%%%%%%%%%%%%%%%%%%%%

%\widetext

%%%%%%%%%%%%%%%%%%%%%%%%%%%%%%%%%%%%%%%%%%%%%%%%%%%%%%%%%%%%%%%%%%%%%%%%%%

Entanglement is one of the most distinct quantum features of many-particle
systems and has only recently started to be studied in a more systematic
way. It provides strong tests of quantum nonlocality \cite{Bell}
and is at the heart of quantum information science with numerous
applications \cite{Bennett,Deutsch,Shor,Deutsch2}. One of the open practical
problems is to identify mechanisms for its robust and controlled generation.
Recently M\o lmer and S\o rensen suggested a scheme to create the $N$%
-particle analog of the GHZ state in an ion-trap without the need for a
precise control over the collective vibrational modes of the ions \cite
{MoelmerPRL99}. Due to Kramers degeneracy \cite{Kramers} in the underlying
nonlinear Hamiltonian
different unitary operations needed to be applied for
even and odd number of particles. The optimum interaction time $T$ scales
linearly with the number of particles and extreme fine
tuning of $T$ is required. As a consequence this method is
highly sensitive to variations of external and internal parameters and 
in the presence of decoherence the
success probability decreases exponentially with $N$.

In the present paper we discuss a robust and efficient method to
create complex entangled states like the $N$-particle analog of the GHZ
state \cite{Greenberger}
\begin{eqnarray}
\frac{1}{\sqrt{2}}\Bigl(\left|a_1 a_2\dots a_N\right\rangle +\left|b_1
b_2\dots b_N\right\rangle\Bigr)  \label{GHZ}
\end{eqnarray}
in a 3-step adiabatic process in a total interaction time which scales
substantially less than linear in the number of particles. The underlying
interaction is the resonant dipole-dipole interaction and the associated
blockade effect in an ensemble of frozen Rydberg atoms studied in 
\cite{Lukin01}. It
was shown in \cite{Lukin01} that the dipole-blockade can be used to generate
any (symmetric) entangled many-particle state by applying a specific
sequence of resonant pulses \cite{Bouchoule01}. A substantial drawback is
however the need of a large number of pulses (which scales linear in $N$)
with well defined pulse area. Hence the method is also highly sensitive to
variations of external and internal parameters and due to the intermediate
excitation of decaying Rydberg levels the success probability decreases
exponentially with $N$.

Following the proposal of Ref.\cite{Lukin01} let us consider 
an ensemble of $N$ atoms with two
lower levels $|a\rangle $ and $|b\rangle $ both coupled to a Rydberg state 
$|r\rangle $ by coherent laser fields with (real) Rabi-frequencies $\Omega
_{1}(t)$ and $\Omega _{2}(t)$ respectively. 
Let us further assume that there are two 
additional Rydberg levels above and below $|r\rangle$ 
with equal energy splitting.  In such a
configuration there is a resonant energy transfer 
between two atoms in Rydberg levels, leading to a symmetric
splitting of the doubly-excited 
states. If the minimum splitting, given by
the atoms of largest separation, exceeds the natural linewidth, resonant
laser excitation into doubly- or higher excited states is
suppressed (dipole-blockade).
In the absence of this 
dipole-blocking, the Hamiltonian is linear in the total spin of the atoms
(SU($2$) symmetry) and it is not possible to create entanglement.

On the other hand for perfect dipole-blockade there is never more than one
excitation in the Rydberg levels. In this limit the effect of the
dipole-dipole interaction can easily be modeled by treating atoms in the
Rydberg state as fermions $(\sigma ,\sigma ^{+})$, while representing atoms
in levels $|a\rangle $ and $|b\rangle $ by bosons $(a,a^{+})$, $(b,b^{+})$.
The presence of the fermionic component breaks the SU($2$) symmetry
and the interaction is no longer linear in the total spin but can 
be described by a multi-mode Jaynes
Cummings Hamiltonian \cite{JC-review}: 
\be
H(t)=\Delta _{1}a^{+}a+\Delta _{2}b^{+}b+\Bigl(\Omega _{1}a+\Omega
_{2}b\Bigr)\sigma ^{+}+h.c. 
\ee
The detunings $\Delta _{i}$ have to be much smaller than the minimum
splitting of the doubly-excited manifold for the blockade-limit to hold. 

The isomorphism to the multi-mode Jaynes Cummings model has a number of
interesting consequences. First of all it simplifies the
analysis by allowing to employ angular momentum techniques. Secondly many
known features of the Jaynes-Cummings dynamics, such as decay and revivals
of oscillations \cite{JC-revivals}, squeezed-state generation, and 
quantum state transfer between different modes \cite{JC-Stenholm} 
can be anticipated in the dipole-blocking system.
 
The blockade of double and higher excitations results in a chainwize
coupling between symmetric collective states as shown in Fig.~1. This
coupling with an odd total number of levels suggests the application of
dark-state Raman adiabatic passage techniques \cite{Vitanov01}. To analyze
adiabatic passage in such a system it is convenient to introduce dark- and
bright-state boson operators 
\begin{eqnarray}
D=a \cos \theta -b\sin \theta,\qquad B=a \sin \theta +b\cos \theta,
\label{dark}
\end{eqnarray}
with $\tan\theta=\Omega_1/\Omega_2$. In terms of these variables the
interaction Hamiltonian reads 
\begin{eqnarray}
H &=& \frac{\Delta_1+\Delta_2}{2} \bigl(D^+ D +B^+ B\bigr)  \nonumber \\
&+&\frac{\Delta_1-\Delta_2}{2} \bigl(D^+ D -B^+ B\bigr)\cos 2\theta
\label{Two-modeJC} \\
&+& \frac{\Delta_1-\Delta_2}{2} \bigl(D^+ B -B^+ D\bigr)\sin 2\theta 
\nonumber \\
&+&\Omega_0\bigl(B\, \sigma^+ +B^+\, \sigma\bigr),  \nonumber
\end{eqnarray}
with $\Omega_0=\sqrt{\Omega_1^2(t)+\Omega_2^2(t)}$. The first two terms are
the free energy of the atoms in the dark and bright states and the third
term describes the coupling between dark and bright states. The last term
shows that only the bright-state component is coupled to the Rydberg levels.

%%%%%%%%%%%%%%%%%%%%%%%%%%%%%%%%%%%%%%%%%%%%%%%%%%%%%%%%%%%%%%%%%%%%%%%%

\ 

\begin{figure}[ht]
\begin{center}
\epsfig{file=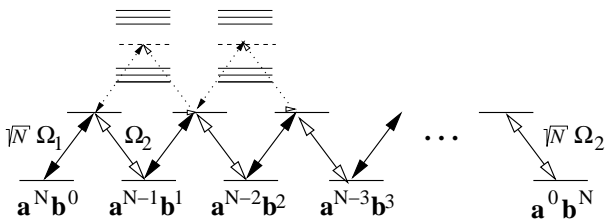,width=8 cm}
\vspace*{4ex}
\caption{Coupling scheme of collective $N$-atom states in limit of dipole
blockade, shown here for $\Delta_1=\Delta_2=0$. Individual atoms have two
lower states $|a\rangle$ and $|b\rangle$ coupled to Rydberg state $|r\rangle$
with Rabi-frequencies $\Omega_1$ and $\Omega_2$ respectively. $|{\bf a}^{N-m}
{\bf b}^m\rangle$ denotes symmetric superposition of $N-m$ atoms in state $
|a\rangle $ and $m$ atoms in state $|b\rangle$. }
\end{center}
\label{chain}
\end{figure}

%%%%%%%%%%%%%%%%%%%%%%%%%%%%%%%%%%%%%%%%%%%%%%%%%%%%%%%%%%%%%%%%%%%%%%%%%%%%

Under conditions of two-photon resonance, i.e. $\Delta _{1}=\Delta _{2}$ the
dark-state subspace decouples from the remaining system. Its dynamics has
however again SU($2$) symmetry and factorized states remain factorized.
Hence {\it dark-state} adiabatic transfer is not suitable for entanglement
generation.
This result can easily be understood physically. Since the
dark state does not contain any excited-state population, the presence of
dipole-dipole interactions and the resulting dipole blockade are irrelevant.
 Nevertheless, as will be shown in the following,
 adiabatic techniques can be used to create entanglement
if other than the zero-eigenvalue state are involved.

To this end we
consider here a situation opposite to the two-photon resonance
when $\Delta =\Delta _{1}=-\Delta_{2}$. We first discuss the case of a
substantial delay between the two pulses $\Omega _{1}$ and $\Omega_2$ such
that the coupling between the dark and bright components, which is
proportional to $\Delta \sin 2\theta$ is negligible. In this approximation
the Hamiltonian (\ref{Two-modeJC}) can be expressed in the simple form 
\begin{equation}
{H}=\frac{\Delta }{2}\sigma _{z}\cos 2\theta +\Omega _{0}\left( B\sigma
^{+}+B^{\dagger }\sigma ^{-}\right) ,  \label{approx_Schrodinger}
\end{equation}
where the irrelevant constant term $N\Delta/2 \cos 2\theta $ was omitted.
The corresponding Schr\"odinger equation can be solved analytically in the
adiabatic limit, i.e. when the mixing angle $\theta(t)$ changes sufficiently
slowly in time.

To obtain a convenient closed form of the solution we introduce angular
momentum operators  $J_{1} =
( \sigma ^{+}B+\sigma ^{-}B^{\dagger })/2\sqrt{M}, 
J_{2} = i( \sigma ^{+}B-\sigma ^{-}B^{\dagger })/2\sqrt{M}$, and $
J_{3} =\sigma _{z}/2$,
where $M=B^{\dagger }B+\sigma ^{+}\sigma $ is the constant particle number
in the bright-state--Rydberg manifold. In terms of these operators the
Hamiltonian reads
$H= \overline{\Omega} \,{\rm e}^{-i\beta J_{2}}\, J_{3}\, {\rm e}^{i\beta
J_{2}}, $
where $\overline{ \Omega} \equiv \sqrt{\Omega_{0}^{2}(t)+\Delta^{2}\cos^{2}2%
\theta }$, and $\tan \beta (t) =\Omega _{0}(t)/\Delta \cos2\theta$. The
corresponding unitary evolution operator then reads
\begin{equation}
U\left( t\right) ={\rm e}^{-i\beta \left( t\right) J_{2}}\exp \biggl[ %
-iJ_{3}\int\limits_{-\infty }^{t}\bar\Omega \left( t^{\prime }\right) {\rm d}%
t^{\prime }\biggr] {\rm e}^{i\beta \left( -\infty \right) J_{2}}.
\label{final solution}
\end{equation}

Let us now consider the case of all $N$ atoms being initially in $|a\rangle $%
. If an intuitive pulse sequence is applied, i.e. if $\Omega _{1}$ is
switched on and off before $\Omega _{2}$ one has $\beta : 0 \rightarrow \pi$
and the systems starts from a bright state 
\begin{equation}
|\Psi_0\rangle=|{\bf a}^N\rangle \equiv\frac{1}{\sqrt{N!}}\left( a^{\dagger
}\right) ^{N}\left| 0\right\rangle =\frac{1}{\sqrt{N!}}\left( B^{\dagger
}\right) ^{N}\left| 0\right\rangle ,  \label{initial-state}
\end{equation}
where $|{\bf a}^N\rangle$ denotes the  collective state with all
atoms being in level $|a\rangle$. The unitary evolution 
operator then reads
\begin{eqnarray}
W &=& -2iJ_{2}\, \exp \biggl[ -iJ_{3} \int\limits_{-\infty }^{+\infty }\!\!%
{\rm d}t\,\beta(t) \biggr].
\end{eqnarray}
One-time application of $W$ generates a symmetric collective
state containing a single Rydberg excitation and all other atoms
are in $|b\rangle$: 
\begin{eqnarray}
\left| \Psi_1 \right\rangle &=&W\,\left| \Psi_0 \right\rangle
= \frac{1}{\sqrt{\left( N-1\right) !}}
\left( B^{\dagger }\right) ^{N-1}\sigma ^{+}\left| 0\right\rangle  \nonumber
\\
&=& \frac{1}{\sqrt{\left( N-1\right) !}}\left( b^{\dagger }\right)
^{N-1}\sigma ^{+}\left| 0\right\rangle ,  \label{1_W_state}
\end{eqnarray}
corresponding to 
\begin{eqnarray}
|{\bf a}^N\rangle\, \stackrel{W}{\longrightarrow}\, |{\bf b}^{N-1}{\bf r}\rangle.\label{W-a}
\end{eqnarray}
Applying $W$ twice generates the W-state of Ref.\cite{Duer00}
\begin{eqnarray}
|{\bf a}^N\rangle\, \stackrel{W^2}{\longrightarrow}\, |{\bf a}^{N-1}{\bf b}\rangle.\label{W2}
\end{eqnarray}
On the other hand starting from an initial state  
with all atoms in 
$|b\rangle$ corresponds to a pulse sequence in counter-intuitive order
and leads to the transfer
\begin{equation}
|{\bf b}^N\rangle\, 
\stackrel{W}{\longrightarrow}\, |{\bf a}^{N}\rangle.\label{W-b}
\end{equation}

Iterative applications of the {\it same} operator $W$ allows one to reach
any state in the $2N+1$ dimensional manifold of symmetric many-particle
excitations with at most one Rydberg atom. The $W$-operation is based on
adiabatic evolution and is robust against variations of parameters as long
as the condition 
\begin{eqnarray}
\gamma\int_{-\infty}^\infty {\rm d} t\, \frac{\dot\theta^2(t)} {\overline{%
\Omega}^2(t)} \ll 1
\end{eqnarray}
is fulfilled, with $\gamma$ being the decay rate of the Rydberg levels.

Although the application of $W$ leads to an entangled state whose creation
would require many $\pi$-pulses, ${\cal O}(N)$ steps are needed for the
generation of complex states like the $N$-particle analog of the GHZ state (%
\ref{GHZ}). We will now show that a small modification of the $W$ operation
can achieve this goal in very few steps and independent on the number of
particles.

For this we assume that the system is initially in an equal superposition of
atoms being in $|a\rangle $ and the symmetric state containing a single
Rydberg excitation. 
\begin{eqnarray}
|\Psi _{0}^{\prime }\rangle 
=\frac{1}{\sqrt{2}}\Bigl(|{\bf a}^{N}\rangle +|{\bf a}^{N-1}{\bf r}\rangle
\Bigr).
\end{eqnarray}
$|\Psi _{0}^{\prime }\rangle $ can easily be created out of $%
|\Psi _{0}\rangle $ in a robust way e.g. by sweeping the frequency of $%
\Omega _{1}$ through resonance (rapid adiabatic passage) \cite{Vitanov00}.
We now apply the $W$ operation discussed above, however with a smaller time
delay between the two pulses. In this case the dark-bright state coupling in
the Hamiltonian (\ref{Two-modeJC}) proportional to $\sin 2\theta $ needs to
be taken into account. Furthermore it is assumed that $|\Omega
_{0}|\gg |\Delta |$. Under these conditions the Schr\"{o}dinger equation can
no longer be solved analytically. However numerically evaluating the 
equations (for $N$ up to 20), we found the behavior shown in Fig.~2.

The mechanism can qualitatively be understood as
follows: Due to the non-vanishing detuning $\Delta $ and the chosen 
intuitive pulse
order, the state amplitude in $|{\bf a}^{N}\rangle $ undergoes an adiabatic
return process \cite{Vitanov00} and ends up in the same state as it started
from. At the same time the chosen pulse order is counter-intuitive for the
state $|{\bf a}^{N-1}{\bf r}\rangle $ and hence its amplitude undergoes Raman
adiabatic passage to $|{\bf b}^{N-1}{\bf r}\rangle $ through a chain of
successive $V$-type transitions. Since the fields are not in $N$-photon
resonance, it is essential that $|\Omega _{\rm m}|\gg |\Delta |$.

%%%%%%%%%%%%%%%%%%%%%%%%%%%%%%%%%%%%%%%%%%%%%%%%%%%%%%%%%%%%%%%%%%%%%%%%

\ 

\begin{figure}
\begin{center}
\epsfig{file=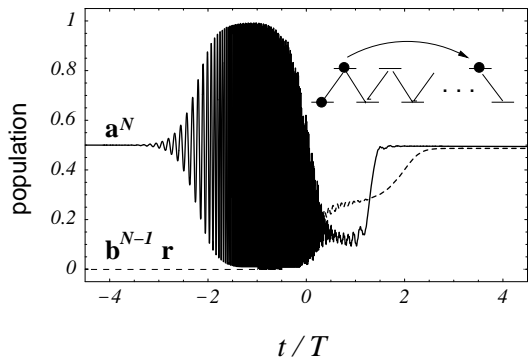,width=7 cm}
\vspace*{4ex}
\caption{Temporal evolution of population in $|{\bf a}^{N}\rangle $
and $|{\bf b}^{N-1}{\bf r}\rangle $ form initial state $|\Psi _{0}^{\prime
}\rangle $ for $N=5$. The laser pulses are Gaussian $%
\Omega _{1,2}\left( t\right) =\Omega_{\rm m}$ $\exp [-\left( t\pm \protect\tau
\right) ^{2}/T^{2}]$, the delay is $\protect\tau =0.5T$ the pulse area $
\Omega_{\rm m} T=125$, and $\Delta T=50$.}
\end{center}
\label{num1}
\end{figure}

%%%%%%%%%%%%%%%%%%%%%%%%%%%%%%%%%%%%%%%%%%%%%%%%%%%%%%%%%%%%%%%%%%%%%%%%%%%%

The amplitude of the state vector in $|{\bf a}^{N}\rangle $ undergoes some
rapid oscillations but returns to the same state for $t\rightarrow \infty $.
At the same time the amplitude in $|{\bf a}^{N-1}{\bf r}\rangle $ is
transferred to $|{\bf b}^{N-1}{\bf r}\rangle $. 
Applying in a third step the inverse of $W$, eqs.(\ref{W-a}) and (\ref{W-b}),
eventually leads to the $N$-particle GHZ state
(\ref{GHZ}). This
corresponds to the overall 3-step adiabatic process: 
\begin{eqnarray}
&&|{\bf a}^{N}\rangle \,\rightarrow \,\frac{1}{\sqrt{2}}\Bigl(|{\bf a}%
^{N}\rangle +|{\bf a}^{N-1}{\bf r}\rangle \Bigr)\,\rightarrow  \\
&&\qquad \frac{1}{\sqrt{2}}\Bigl(|{\bf a}^{N}\rangle +|{\bf b}^{N-1}{\bf r}%
\rangle \Bigr)\,\rightarrow \,\frac{1}{\sqrt{2}}\Bigl(|{\bf b}^{N}\rangle +|%
{\bf a}^{N}\rangle \Bigr).  \nonumber
\end{eqnarray}

The transfer is in all parts robust. It does not depend on the exact pulse
form of $\Omega _{1}$ and $\Omega _{2}$, nor does it require an
extreme control of the delay time $\tau$ or the pulse length $T$.
Furthermore the mechanism works for even and odd numbers of atoms in
the same way. In Fig.~3 we have shown the dependence of the final population
in the states $|{\bf a}^N\rangle$ and $|{\bf b}^{N-1}{\bf r}\rangle$ for $N=5$
as function of pulse delay $\tau$ and pulse area $\Omega_{\rm m} T$. 
It can be seen 
that the mechanism is robust against small variation of the delay
time and -- above some critical limit -- the pulse area. 
It should be mentioned that for very large values of the pulse area 
the populations decrease again, since then 
the term $\Delta \sin 2\theta $ in (\ref{Two-modeJC}) is negligible and
there is a transfer 
$\frac{1}{\sqrt{2}}\bigl
(|a^{N}\rangle +|a^{N-1}r\rangle \bigr) \to 
\frac{1}{\sqrt{2}}\bigl(|b^{N}\rangle +|b^{N-1}r\rangle \bigr).
$

Since all processes are adiabatic and $\Omega_{\rm m}$ 
is limited by the dipole-blockade condition, the question arises how the
time $T$ for generating the GHZ state scales with the number of particles.
From our numerical calculations, shown in Fig.~4, we find $T\sim N^{\alpha }$
with $\alpha< 1$ and decreasing with $N$. 
The numerical calculations for $N=3\dots 16$ indicate
$\alpha \to 2/3$. 
Thus in the presence of decay,
 the success probability decreases less than exponential
with $N$.

%%%%%%%%%%%%%%%%%%%%%%%%%%%%%%%%%%%%%%%%%%%%%%%%%%%%%%%%%%%%%%%%%%%%%%%%

\begin{figure}
\begin{center}
\epsfig{file=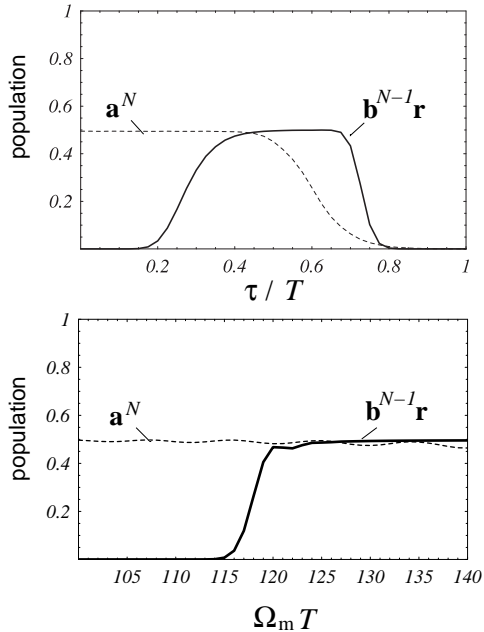,width=6.5 cm}
\vspace*{4ex}
\caption{Final population of states  $|{\bf a}^{N}\rangle $
and $|{\bf b}^{N-1}{\bf r}\rangle $ for conditions of Fig.~2.
{\it top:} as function of delay time $\tau$ for $
\Omega_{\rm m}T=120$, {\it bottom:} as function of $
\Omega_{\rm m}T$ for $\protect\tau =0.5T$.}
\end{center}
\end{figure}

%%%%%%%%%%%%%%%%%%%%%%%%%%%%%%%%%%%%%%%%%%%%%%%%%%%%%%%%%%%%%%%%%%%%%%%%%%%%

%%%%%%%%%%%%%%%%%%%%%%%%%%%%%%%%%%%%%%%%%%%%%%%%%%%%%%%%%%%%%%%%%%%%%%%%

\ 

\begin{figure}[th]
\begin{center}
\epsfig{file=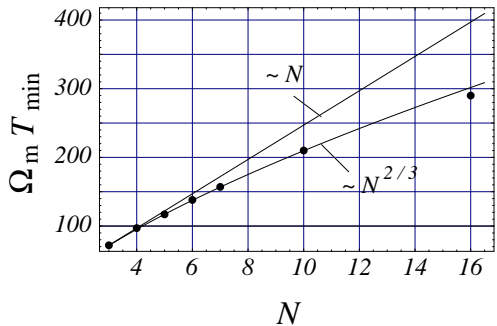,width=6.5 cm}
\vspace*{4ex}
\caption{Minimum pulse area $ \Omega_{\rm m} T_{\rm min}$
to create $GHZ$ state as function of particle number $N$. Dots
represent values from numerical solution of $N$-particle Schr\"odinger equation
for optimized $\tau$.}
\end{center}
\label{num2}
\end{figure}

%%%%%%%%%%%%%%%%%%%%%%%%%%%%%%%%%%%%%%%%%%%%%%%%%%%%%%%%%%%%%%%%%%%%%%%%%%%%

In conclusion, we have proposed an efficient and robust method to generate
complex entanglement structures, such as the $N$-particle GHZ state in a
many-particle system with resonant dipole-dipole interactions. 
The method is robust against variations of parameters since
for all steps adiabatic transfer processes are used. Although dark-state
adiabatic passage is not suitable for entanglement generation, as it does
not involve population of the interacting Rydberg levels, other adiabatic
processes are identified that allow e.g. for the generation of the $N$%
-particle GHZ state (\ref{GHZ}) in three steps. The suggested method avoids
the problem associated with Kramers degeneracy and thus works for even and
odd number of particles. Exact knowledge of the number of particles
is not required, making the method robust against number fluctuations. As
opposed to the proposal of ref.\cite{MoelmerPRL99} no extreme fine tuning of
the interaction time is needed and the minimum interaction time scales only
less than linear with the number of particles. Finally it should be
mentioned that similar ideas can be applied to other
many-particle systems, e.g. ions in a trap.

The work of RU was supported by the DFG under contract number FL 210/10 and
the Alexander von Humboldt Foundation. We thank K. Bergmann and B.W. Shore
for helpful discussions.

%%%%%%%%%%%%%%%%%%%%%%%%%%%%%%%%%%%%%%%%%%%%%%%%%%%%%%%%%%%%%%%%%%%%%%%%%%%

\end{document}